\documentclass{caosp309}
\usepackage{graphicx}

\usepackage{natbib}
\articleNo{?}
\pubyear{2025}
\volume{?}
\volnumber{?}
\firstpage{1}
\received{December 12, 2024}
\accepted{January 11, 2025}

\def\BibTeX{{\rm B\kern-.05em{\sc i\kern-.025em b}\kern-.08em
             T\kern-.1667em\lower.7ex\hbox{E}\kern-.125emX}}

\begin{document}

%%%%%%%%%%%%%%%%%%%%%%%%%%%%%%%%%%%%%%%%%%%%%%%%%%%%%%%%%%%%%%%%%%%%%%%%%%%%%
%              R U N N I N G   P A G E   H E A D I N G S                     
% Odd page headings (except for the title page) are produced automatically
% and contain the title. If, and only if, the title of your article is too
% long the running head is omitted in the printout; you can make your own
% running title by using the \htitle command, putting the shortened title
% between the curly brackets. \htitle should also be used when the
% subtitle is present: \htitle offers you a way how to include it into the
% headings. If you wish to see how it works simply remove the % sign from
% the beginning of that line.
%
% Unlike the \htitle command, the \hauthor command is compulsory. It is
% used to produce even page headings and contains the names of the authors
% of an article.  All authors must be listed here, if possible. When
% authors' list is too long, you can abbreviate it by using "{\it et
% al.}". Authors' names are given in the form: initial(s) of the author's
% first name and surname. Authors are separated by a "," (comma) sign and
% the last one by "and".
%%%%%%%%%%%%%%%%%%%%%%%%%%%%%%%%%%%%%%%%%%%%%%%%%%%%%%%%%%%%%%%%%%%%%%%%%%%%%
%\htitle{A note to comet ejection process ...}
\hauthor{B. Pilecki}
%\hauthor{L.\,Neslu\v{s}an {\it et al.}}

%%%%%%%%%%%%%%%%%%%%%%%%%%%%%%%%%%%%%%%%%%%%%%%%%%%%%%%%%%%%%%%%%%%%%%%%%%%%%
%                       T I T L E                                            
% Capital letters in the title are only used at the beginning of the
% names. Don`t end the title by a "." (dot)
%%%%%%%%%%%%%%%%%%%%%%%%%%%%%%%%%%%%%%%%%%%%%%%%%%%%%%%%%%%%%%%%%%%%%%%%%%%%%
\title{Cepheids in spectroscopic binary systems - current status and recent discoveries}

%%%%%%%%%%%%%%%%%%%%%%%%%%%%%%%%%%%%%%%%%%%%%%%%%%%%%%%%%%%%%%%%%%%%%%%%%%%%%
%                       S U B T I T L E                                      
% You can use the subtitle, with the command \subtitle similar to the
% \title command.
%%%%%%%%%%%%%%%%%%%%%%%%%%%%%%%%%%%%%%%%%%%%%%%%%%%%%%%%%%%%%%%%%%%%%%%%%%%%%

%%%%%%%%%%%%%%%%%%%%%%%%%%%%%%%%%%%%%%%%%%%%%%%%%%%%%%%%%%%%%%%%%%%%%%%%%%%%%
%                   A U T H O R  N A M E S                                   
% Authors' names are separated by the \and command and their institutes
% are assigned by the \inst{n} command. 
% If all authors belong to just one institute, it is not needed/desired
% to use the \inst command.
%
% Author can indicate her/his ORCID (https://orcid.org/) identifier using
% the command \orcid. It will not appear in the LaTeX output but will be
% sent to the ADS database.
%
% When the name contains "Slovak" letters L,d,t,l with a caron, use an
% a new \softl, etc. command (examples given in the last table of
% this document) to produce typographically correct accented characters.
%%%%%%%%%%%%%%%%%%%%%%%%%%%%%%%%%%%%%%%%%%%%%%%%%%%%%%%%%%%%%%%%%%%%%%%%%%%%%
\author{
        B.\,Pilecki\orcid{0000-0003-3861-8124}
       }

%%%%%%%%%%%%%%%%%%%%%%%%%%%%%%%%%%%%%%%%%%%%%%%%%%%%%%%%%%%%%%%%%%%%%%%%%%%%%
%           I N S T I T U T E S'  A D D R E S S E S                          
% The affiliation of authors is generated by the \institute command, the
% \and command being again used to separate individual addresses.
% The following commands may be used for the following three institutes:   
%               \lomnica        for      AsU SAV, Tatranska Lomnica          
%               \blava          for      AsU SAV, Bratislava                 
%               \ondrejov       for      AsU CAV, Ondrejov                   
%
% The given postal address must be complete in order to facilitate our
% editorial work. Moreover, you can add your e-mail address, using the
% \email command.
%%%%%%%%%%%%%%%%%%%%%%%%%%%%%%%%%%%%%%%%%%%%%%%%%%%%%%%%%%%%%%%%%%%%%%%%%%%%%
\institute{
           Centrum Astronomiczne im. M. Kopernika PAN, 
           Warsaw, Poland\\ \email{pilecki@camk.edu.pl}
          }

%%%%%%%%%%%%%%%%%%%%%%%%%%%%%%%%%%%%%%%%%%%%%%%%%%%%%%%%%%%%%%%%%%%%%%%%%%%%%
%                        D A T E / R E C E I V E D                          
% Date inserted here will be the date when your paper was received The
% format is: month (not abbreviated), day, year.
%%%%%%%%%%%%%%%%%%%%%%%%%%%%%%%%%%%%%%%%%%%%%%%%%%%%%%%%%%%%%%%%%%%%%%%%%%%%%
\date{December 5, 2024}
%\date{March 10, 2003}

%%%%%%%%%%%%%%%%%%%%%%%%%%%%%%%%%%%%%%%%%%%%%%%%%%%%%%%%%%%%%%%%%%%%%%%%%%%%%
%                        M A K E T I T L E
% The beginning part (title, author(s), etc.) of your article must be
% closed by the \maketitle command.
%%%%%%%%%%%%%%%%%%%%%%%%%%%%%%%%%%%%%%%%%%%%%%%%%%%%%%%%%%%%%%%%%%%%%%%%%%%%%
\maketitle

%%%%%%%%%%%%%%%%%%%%%%%%%%%%%%%%%%%%%%%%%%%%%%%%%%%%%%%%%%%%%%%%%%%%%%%%%%%%%
%                        A B S T R A C T,  K E Y W O R D S                   
% Here it is shown how to write an abstract.  Keywords should be placed
% within the "abstract" environment using the command \keywords and they
% should be selected from the thesaurus from Astron.  Astrophys.
% Abstracts. They must be separated from each other by -- (two dashes).
%%%%%%%%%%%%%%%%%%%%%%%%%%%%%%%%%%%%%%%%%%%%%%%%%%%%%%%%%%%%%%%%%%%%%%%%%%%%%
\begin{abstract}
We present a summary of the current knowledge about Cepheids in binary systems. We focus on the most recent findings and discoveries, such as the highly increasing number of confirmed and candidate spectroscopic binary Cepheids and the progress in determining their physical parameters. This includes new and newly analyzed binary Cepheids in the Milky Way and Magellanic Clouds. We will provide an update on the project to increase the number of the most valuable Cepheids in double-lined binary (SB2) systems from six to more than 100. To date, we have confirmed 60 SB2 systems, including detecting a significant orbital motion for 37. We identified systems with orbital periods up to five times shorter than the shortest period reported before and systems with mass ratios significantly different from unity (suggesting past binary interactions, including merger events). Both features are essential to understanding how multiplicity affects the formation and destruction of Cepheid progenitors and how this influences global Cepheid properties. We will also present nine new systems composed of two Cepheids. Only one such double Cepheid system was known before.
\end{abstract}

%%%%%%%%%%%%%%%%%%%%%%%%%%%%%%%%%%%%%%%%%%%%%%%%%%%%%%%%%%%%%%%%%%%%%%%%%%%%%
%                       S E C T I O N I N G                                  
% Any section starts with the command \section as shown below, with the
% title in Initial Capitals and lowercase only. Do not number the sections
% - let LaTeX do that for you - and do not end them by a "." (dot).
%
% The (sub)section titles are typeset in boldface; so, if working in the
% mathematics mode in (sub)section titles, you must use \boldmath and 
% enclose it into curly brackets, e.g. "{\bolmath $R^{2}$}".
%%%%%%%%%%%%%%%%%%%%%%%%%%%%%%%%%%%%%%%%%%%%%%%%%%%%%%%%%%%%%%%%%%%%%%%%%%%%%
\section{Introduction}
\label{sec:intro}

Classical Cepheids (hereafter also Cepheids) are crucial for various fields of astronomy, including stellar oscillations and the evolution of intermediate and massive stars, and have an enormous influence on modern cosmology \citep[see, e.g., the review of][]{Bono_2024_Review_Cepheids}.  Since the discovery of the relationship between their pulsation period and luminosity (the Leavitt Law, \citealt{1912HarCi.173....1L}), the Cepheids have been extensively used to measure distances in the Universe \citep{Anderson_2024book_Cepheid_dist_H0}. The recent local Hubble constant
determination accurate to 1.4\% \citep{Riess_2022_Hubble_1.4}, which
shows a significant discrepancy with the value inferred from {\sl Planck} data, depends sensitively on this period--luminosity (P--L) relation.

Theoretical studies of Cepheids are quite advanced (e.g.,
\citealt{Bono_1999_Cep_TheoryII,Bono_2005_ApJ_CEP_puls_model_IV,Valle_2009_AA_theo_pred_CC_puls}). We know they are radially pulsating evolved intermediate and high-mass giants and supergiants, mostly located in a well-defined position on the helium-burning loop (called the {\em blue loop}). A small fraction of Cepheids is expected to pulsate during their first crossing through the instability strip on their way to the red giant branch.
Empirical relations of different types are also often obtained and applied \citep{Soszynski_2008_LMC_Cepheids, Storm_2011_pfac_AA.534.94, Riess_2019_Hubble_1.9, Espinoza_2024_IS_LMC, Pilecki_2024_fundamentalization}. However, our knowledge about the physical properties of Cepheids is still limited. Theory predicts masses of Cepheids in the range of 3--13 M$_\odot$, but their measured masses clump between 3.6 M$_\odot$ and 5 M$_\odot$, with only one higher but uncertain value of 6.7 M$_\odot$ \citep{allcep_pilecki_2018, Evans_2018_V350_Sgr_mass, Evans_2024_AW_Per_SB1}. This makes the Cepheid mass--luminosity relation poorly constrained \citep{Anderson_2016_Rotation}. Moreover, the blue loops predicted by current evolution theory are too short to explain the existence of low-mass, short-period Cepheids \citep[see, e.g.,][]{Ripepi_2022_MNRAS_VMC_LMC_CEPs, Espinoza_2024_IS_LMC}.

In the last decade, the expected fraction of classical Cepheids in multiple systems has grown to about 80\% \citep{Kervella_2019_Multiplicity}, indicating that we cannot ignore their multiplicity in our studies. The light of a companion directly affects the observed brightness of Cepheids, but the tidal deformation, mass transfer, and merger origin may also impact their intrinsic brightness. The binary interactions, changing the stellar rotation, chemical composition, and mass, may affect the evolution of Cepheids and influence their pulsation characteristics. 

There are almost 15000 known classical Cepheids \citep{Pietruk_2021_CEP_num_MW_total}, mainly in the MW \citep[and references therein]{Pojmanski_2005_ASAS_cat5, Soszynski_2020AcA_upd_MW_ceps} and the Magellanic Clouds \citep{Soszynski_2017AcA_OCVS_MC_Cep}. Although 80\% of these Cepheids are expected to be members of binary systems, only about 0.1\% of them have been found in eclipsing systems, and 0.2\% in double-lined spectroscopic binaries (SB2), while both these features provide indispensable information about the physical properties of Cepheids.

Here, we summarize the current state regarding single and double Cepheids in binary systems. We will focus mainly on those that were or can be used to determine the physical parameters of Cepheids, thus primarily focusing on eclipsing and SB2 Cepheids.

%%%%%%%%%%%%%%%%%%%%%%%%%%%%%%%%%%%%%%%%%%%%%%%%%%%%%%%%%%%%%%%%%%%%%%%%%%%%%
\section{Binarity of Cepheids}

Since the discovery of the binarity of Polaris in 1929 \citep{Moore_1929_Bin_Polaris}, many Cepheids in binary systems in the Milky Way were identified. In the updated catalog of \citet{Szabados_2003_BinCep_MW}\footnote{https://webarch.konkoly.hu/cep/intro.html; last update in 2019}, there are 171 Cepheids that are either binary or suspected for binarity. For 31 of them, orbital elements are also provided.
In the recent study based on a large set of new spectroscopic data from the VELOCE project \citep{Anderson_2024_VELOCE}, 32 new Cepheids in single-lined spectroscopic binary (SB1) systems were found, and orbital solutions for 18 of them were provided \citep{Shetye_2024_MW_SB1_Ceps_VELOCE}. 
Unfortunately, all Cepheids from these two samples have main sequence companions that are hardly seen in the spectra. None of these Cepheids were found in an eclipsing system. 

In the Milky Way, there are only two known binary Cepheids without a main sequence companion, for which lines of both components can be easily seen and analyzed. Each comprises two Cepheids, i.e., stars practically in the same evolutionary phase. More information on them is given in Section~\ref{sec:bind}. There is also one candidate for an eclipsing binary Cepheid \citep{Soszynski_2020AcA_upd_MW_ceps}, most probably of SB1 type.

In the analysis of Cepheid variables in the OGLE project \citep{Soszynski_2017AcA_OCVS_MC_Cep}, six candidates for eclipsing binary Cepheids were identified in the Large Magellanic Cloud (LMC) galaxy. Five have already been confirmed as SB2 and analyzed (see Section~\ref{sec:lmc_ecl}). In the Small Magellanic Cloud (SMC) galaxy, there are five candidates for eclipsing binary Cepheids, most probably of SB1 type, of which none has been spectroscopically confirmed yet. However, the light-travel time effect was detected in the photometric time series for one of them \citep{Rajeev_2024_OC_BINCEP}. A nice summary of binary candidates in the Magellanic Clouds is provided in \citet{Szabados_2012_BinCep_MC}. Recently, a new method of detecting SB2 Cepheids was developed that increased the number of such systems by an order of magnitude (see Section~\ref{sec:lmc_newsb2}).

\subsection{Masses of Milky Way Cepheids}
\label{sec:mw_mass}

Almost all binary Cepheids in the Milky Way are single-lined spectroscopic binaries (SB1), making mass estimates for their components very uncertain, with typical accuracies of 10-20\%. Such estimates were done for seven SB1 Cepheids \citep[see,][and references therein]{Evans_2018_V350_Sgr_mass, Evans_2024_AW_Per_SB1}. This is because most binary Cepheids have
an early-type main-sequence companion (exhibiting few and broad lines), typically 2--5 mag fainter in the $V$ band, making it extremely hard to determine its velocity and, thus, the mass of the Cepheid.

However, with considerable effort, more accurate mass measurements could be performed for a few Milky Way Cepheids closest to us. This became possible with the recent improvement in interferometric observations complemented with additional spectroscopic observations from space in far-UV, where companion lines are detectable and radial velocities (RVs) can be measured. However, such spectra are much harder to obtain, which results in the companion's RVs being generally calculated from a low number of spectra. These RVs are also measured separately from the Cepheid RVs, which introduces an extra parameter in the model and increases the measurement uncertainty. These Cepheids are V1334 Cyg \citep{Gallenne_2018_V1334Cyg_Cep}, Polaris \citep{Evans_2024_Polaris_mass}, and SU Cyg \citep{Gallene_2024_SU_Cyg_mass}.

\subsection{Eclipsing binary LMC Cepheids}
\label{sec:lmc_ecl}

Analyzing Cepheids in double-lined eclipsing binaries is the only way to obtain a full physical parametrization of these important stars.  Our spectroscopic observations proved the binarity of five eclipsing Cepheid candidates in the LMC, and their further study brought a wealth of data regarding properties of classical Cepheids, with accuracies of 0.5-2\% for the most interesting Cepheid masses and radii \citep[e.g.,][]{cep227nature2010, cep227mnras2013, cep2532apj2015, allcep_pilecki_2018}. This study also identified several interesting cases, including a rare system composed of two Cepheids, OGLE-LMC-CEP-1718 \citep{cep1718apj2014}. As all these eclipsing Cepheids were accompanied by other giants (in a similar stage of evolution), the measured mass ratios mainly were close to unity, but surprisingly, for one system, significantly different masses ($M_2/M_1$$\sim$0.7) were found. This led to a conclusion that the system (OGLE-LMC-CEP-1812; \citealt{cep1812apj2011}) was a triple before, and the Cepheid is a product of a merger \citep{Neilson2015_cep1812_merger}. It is estimated that even 30\% of Cepheids may be such a product \citep{Sana_2012_BinInteractions}. For the summary, please see \citet{allcep_pilecki_2018}.

\section{Binary Double (BIND) Cepheids}
\label{sec:bind}

Binary systems composed of two Cepheids are crucial for understanding the structure and evolution of these pulsating stars. Having two Cepheids in one system puts strong constraints on models as their ages have to be equal and physical parameters similar and related to their pulsation periods. We can also assume that both Cepheids were born with the same chemical composition. Moreover, the study of OGLE-LMC-CEP-1718 showed that the more evolutionary advanced component is slightly less massive \citep{allcep_pilecki_2018}, which may provide vital information for the ongoing discussion on the origin of the Cepheid mass discrepancy problem \citep{Cassisi_Salaris_2011_ApJ, Anderson_2016_Rotation}.

Although the first visually unresolved pairs of Cepheids (dubbed double Cepheids\footnote{Two Cepheids identified at the same position in the sky (for the used resolution), but not necessarily gravitationally bound.}) were identified almost 30 years ago \citep{Alcock_1995_double_cepheids}, for a long time, only one was spectroscopically confirmed as a binary system and analyzed, which made any statistical analysis impossible. A more extensive set of such stars occupying a wider parameter space would provide the opportunity to gain valuable insight into the pulsation and evolution of Cepheids through a comparative analysis of the differences between the components. All models would have to simultaneously predict correct mass and pulsation period ratios of both Cepheids (together with other observables) for several systems composed of the same age stars with almost the same initial composition.

\begin{figure}
    \centering
    \includegraphics[width=0.32\linewidth]{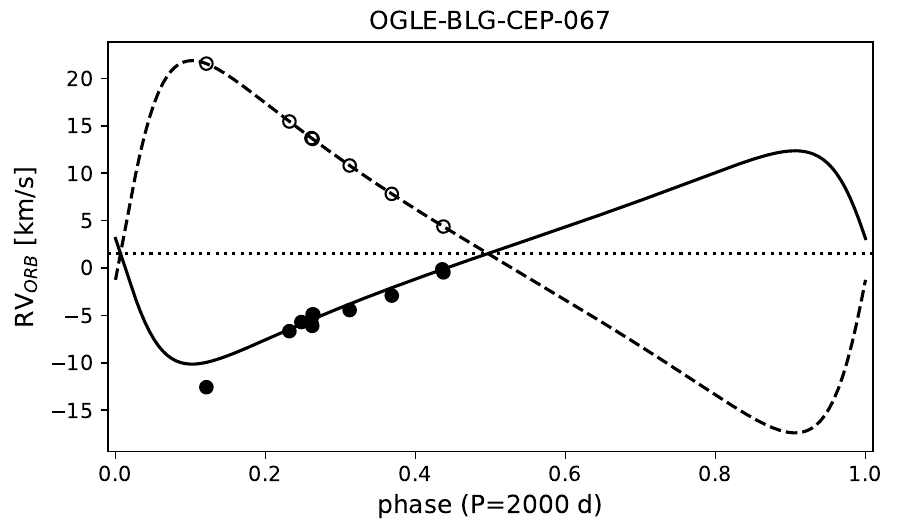}
    \includegraphics[width=0.32\linewidth]{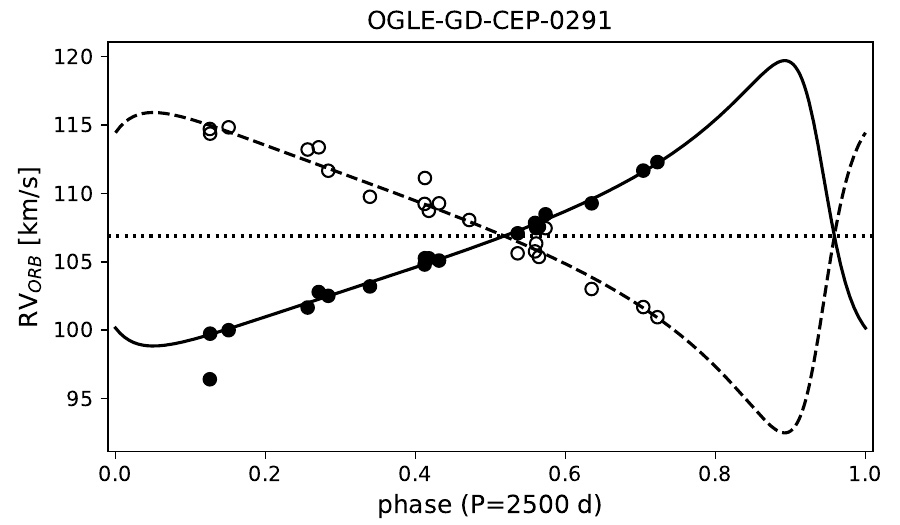}
    \includegraphics[width=0.32\linewidth]{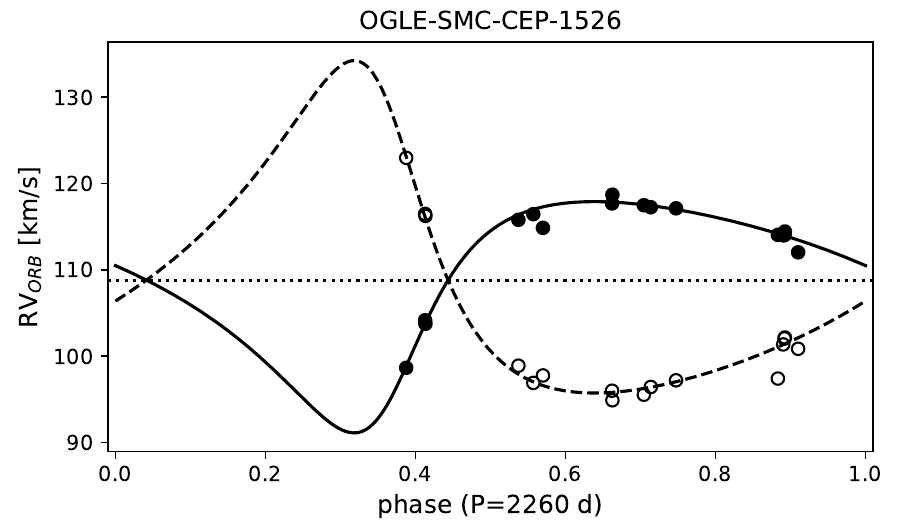}
    \includegraphics[width=0.32\linewidth]{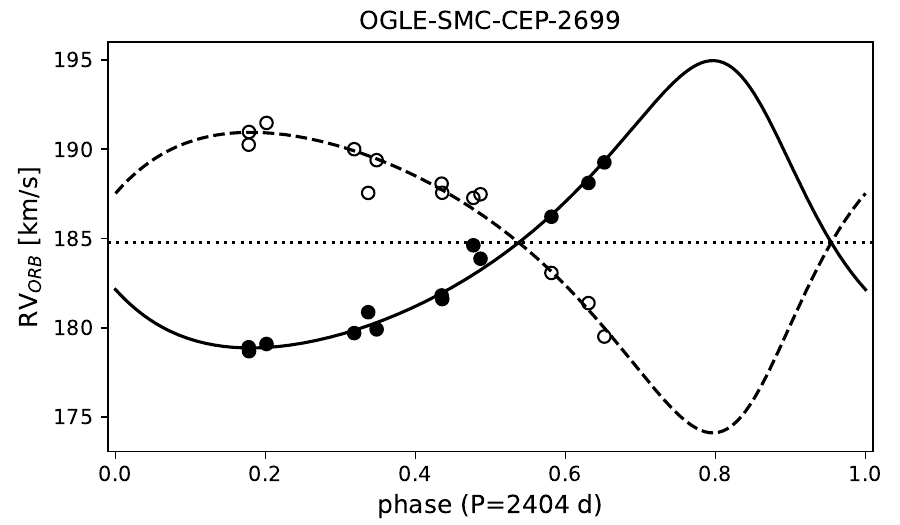}
    \includegraphics[width=0.32\linewidth]{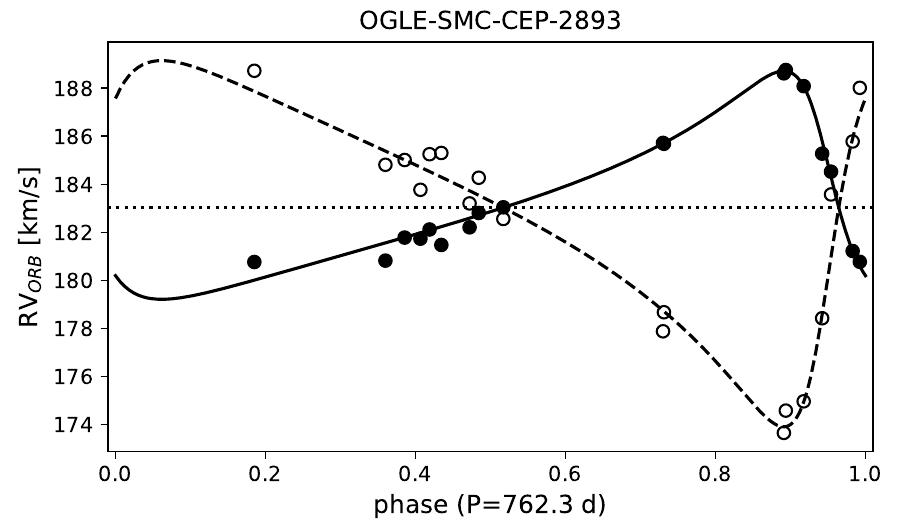}
    \includegraphics[width=0.32\linewidth]{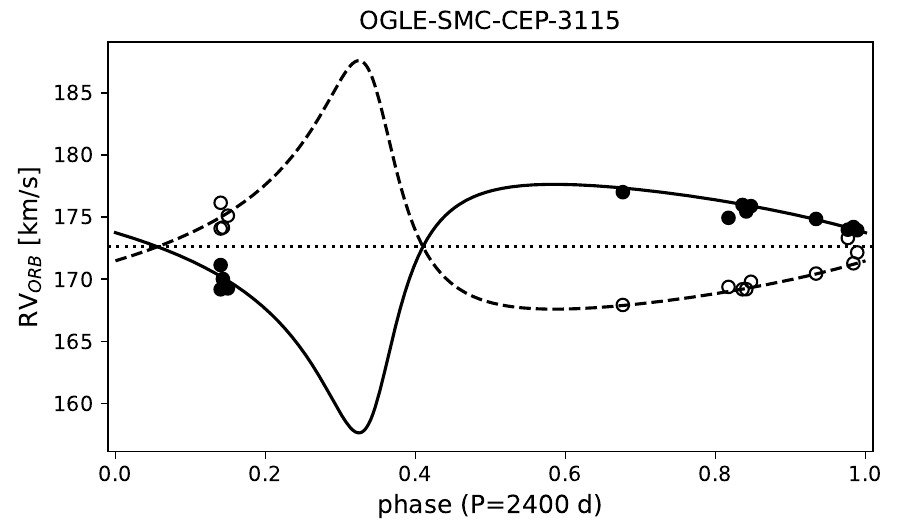}
    \includegraphics[width=0.32\linewidth]{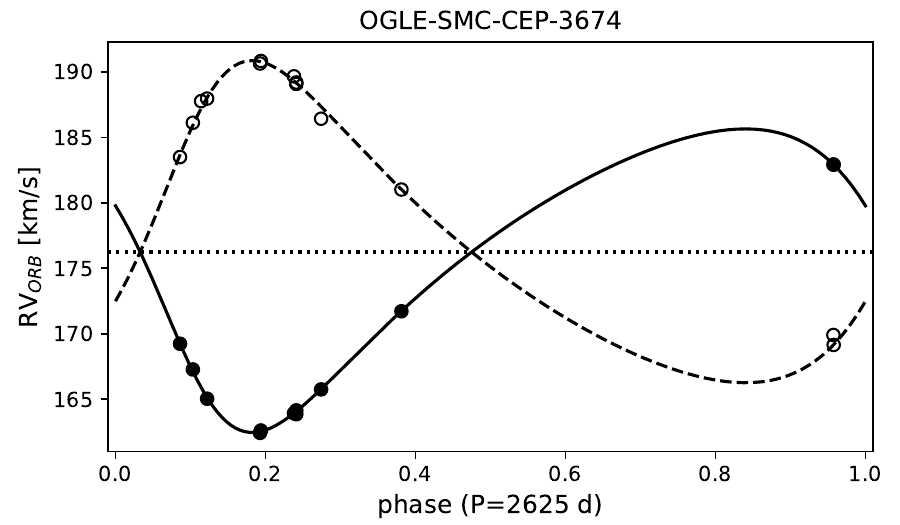}
    \includegraphics[width=0.32\linewidth]{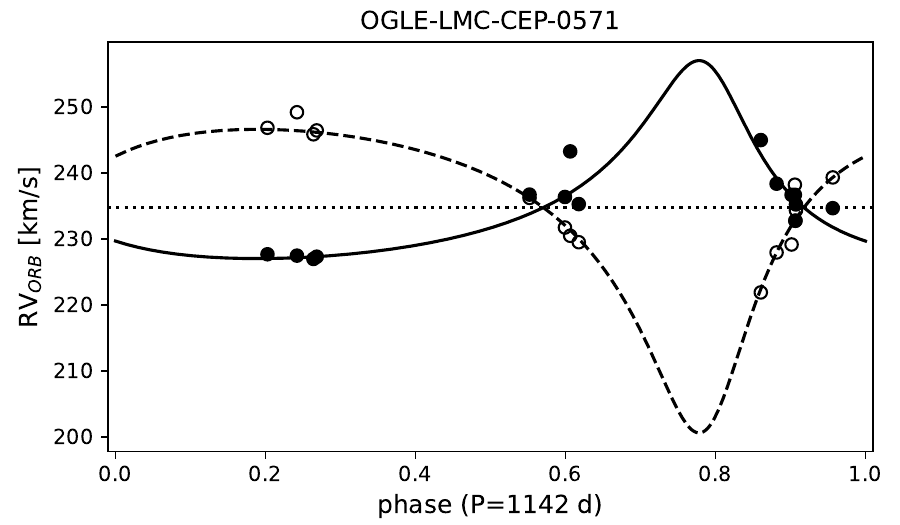}
    \includegraphics[width=0.32\linewidth]{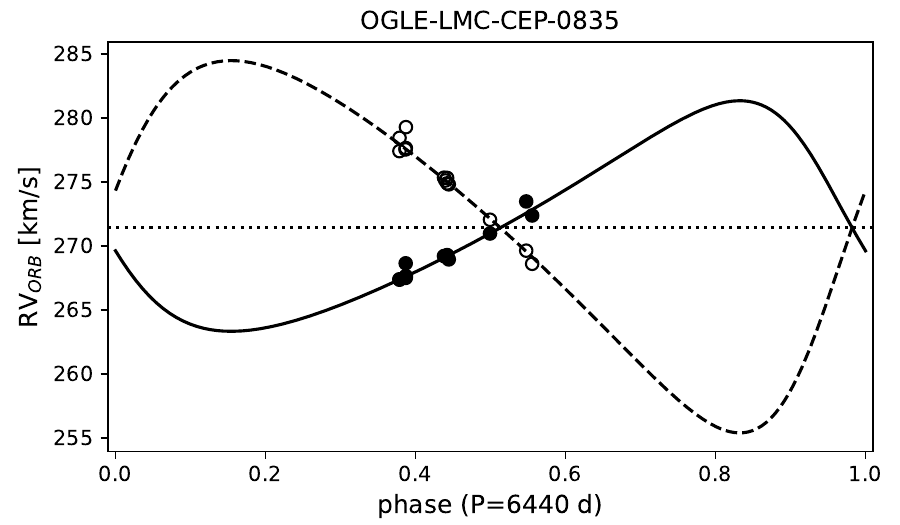}
    \caption{Orbital RV curves for nine new BIND Cepheids. Only for OGLE-SMC-CEP-2893 and OGLE-LMC-CEP-0571 do the orbital periods come from an orbital solution. For three SMC  systems (1526, 2699, 3674) and one in the LMC (0835), the period is taken from the LTTE model, while for the rest, it is set to match the data and not to produce unphysical results.}
    \label{fig:bind}
\end{figure}

Recently, we showed the first results of our observing program of nine candidates for new genuine binary double (BIND) Cepheids, of which two are in the LMC, five in the SMC, and two in the Milky Way. Our spectroscopic observations proved the binarity of all of them \citep{Pilecki_2024_cepgiants2}. The observations were challenging as some systems show very low orbital amplitudes, meaning very low inclinations or long orbital periods. Long orbital periods imply we could not yet cover complete orbital cycles with spectroscopic observations. Fortunately, in some cases, orbital periods could be determined by analyzing the light-travel time effect (LTTE) in the available photometric data, which helped constrain the orbital solutions.
The orbital radial velocity curves after subtracting pulsational variability are shown in Fig.~\ref{fig:bind}. For half of the sample, the ratios of pulsation periods of the components are exceptionally low, which may suggest binary interactions, such as mass transfer or merger events, in their past. This is also proved by a few preliminary mass ratios far from unity. This sample, together with OGLE-LMC-CEP-1812 mentioned above and OGLE-LMC-CEP-1347 described in Section~\ref{sec:lmc_newsb2}, indicate that a significant fraction of Cepheids may be of binary origin. More details about BIND Cepheids can be found in \citet{Pilecki_2024_cepgiants2}.

\section{New SB2 systems and candidates}
\label{sec:lmc_newsb2}

Finding new binary Cepheids, especially in SB2 systems, is crucial for improving our knowledge about these stars. However, most of them have been evading detection for a long time. One good source for binary Cepheids was to look for eclipses, but this solution has three severe limitations. First, such systems are not numerous because a particular orbital orientation is needed for eclipses to occur. Second, according to current statistics, only about half of eclipsing systems involving Cepheids are double-lined, and third, the best places to look for them (the Galactic disk and Magellanic Clouds) have already been examined, and we do not expect many more such binaries to be found there. Therefore, we decided to look for another possible source that had not yet been considered and remained unexplored.

\begin{figure}
    \begin{center}
        \includegraphics[width=0.49\textwidth]{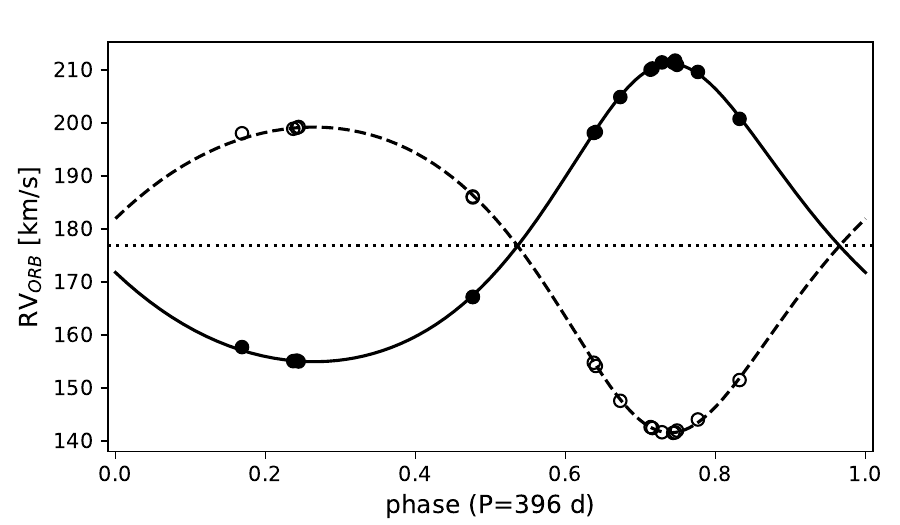}
        \includegraphics[width=0.49\textwidth]{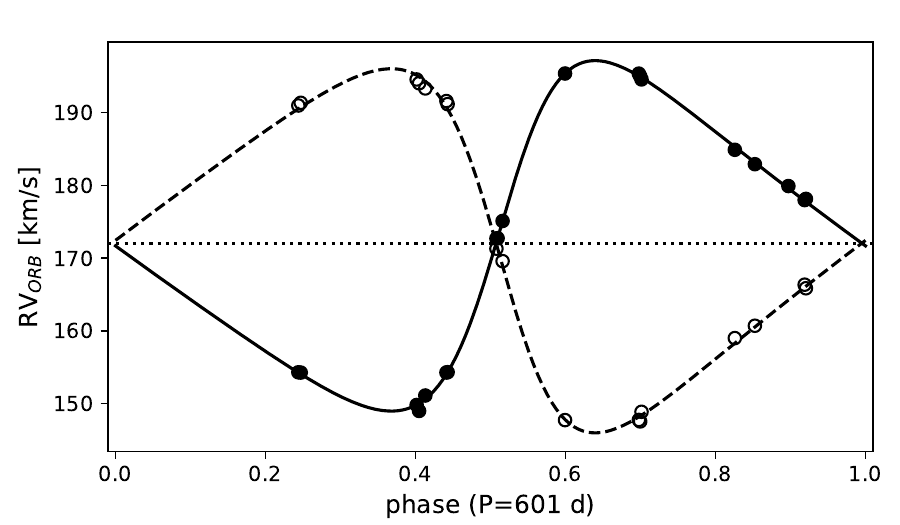}
        \includegraphics[width=0.49\textwidth]{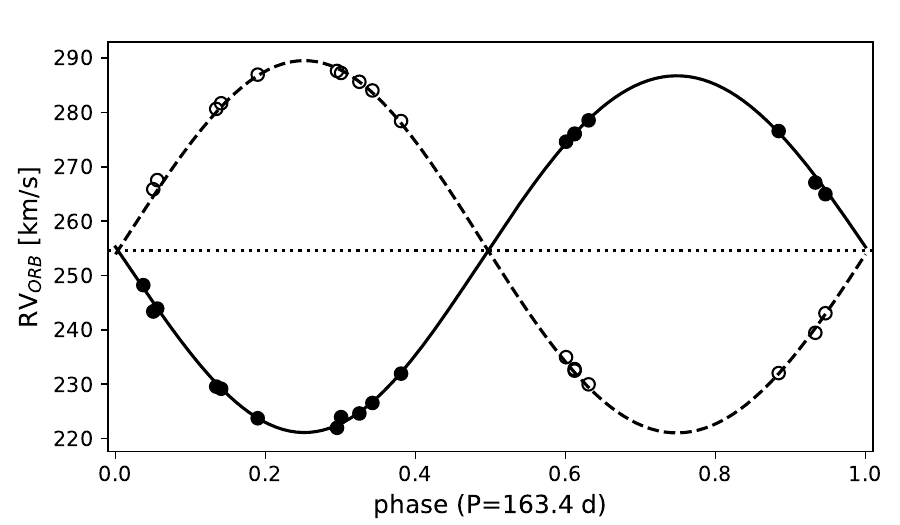}
        \includegraphics[width=0.49\textwidth]{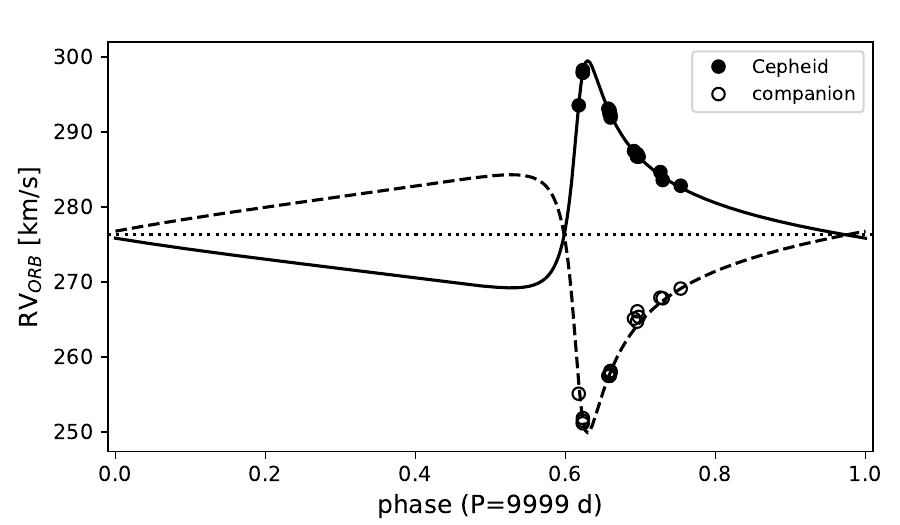}
    \end{center}
\caption{Preliminary orbits for four example double-lined binary Cepheids. The orbital period is not yet constrained for the last one.}
\label{fig:sb2ceps}
\end{figure}

In \citet{cepgiant1_2021}, we proposed a novel method of identification of Cepheids in SB2 systems according to which Cepheids that are excessively bright for their periods, have similar or redder colors, and have lower than average pulsation amplitudes are strongly suspected of forming binaries. There are about 100 such candidates in the Magellanic Clouds.
Using the first collected data for a limited sample, we showed that this method is about 95\% efficient, confirming SB2 status for 17 out of 18 analyzed Cepheids in the LMC galaxy \citep{cepgiant1_2021,cepgiant1pta_2022}. Our current sample of SB2 Cepheids in the LMC, SMC, and Milky Way consists of 60 objects, ten times more than what was known before the project. For 37 of them, an anticorrelated orbital motion of both Cepheid and the companion was detected. This is the ultimate proof that they are gravitationally bound and indicates the orbital periods are relatively short. Preliminary orbital solutions were obtained for 24, with periods of up to about seven years (for the longest periods with the help of the LTTE analysis). Three of them and one for which the orbit is not yet fully constrained are presented in Fig.~\ref{fig:sb2ceps}.

\begin{figure}
    \begin{center}
        \includegraphics[width=0.57\textwidth]{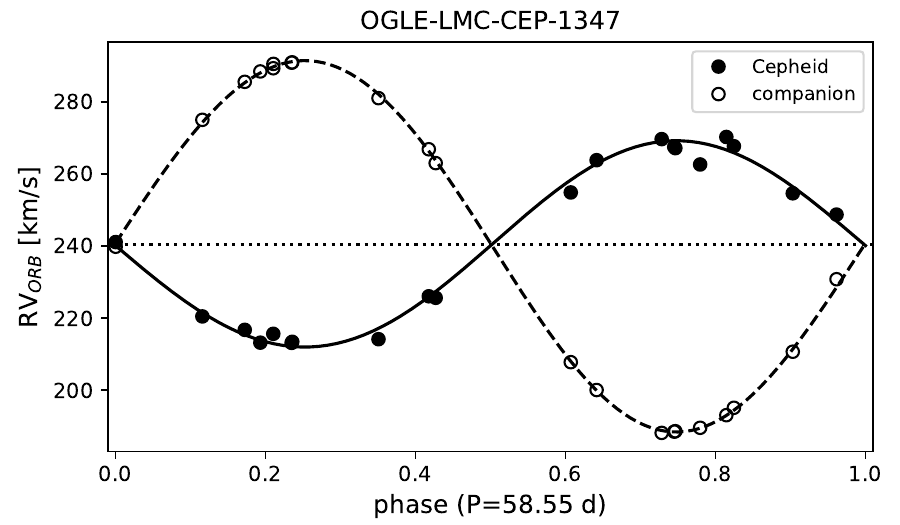}
        \includegraphics[width=0.42\textwidth]{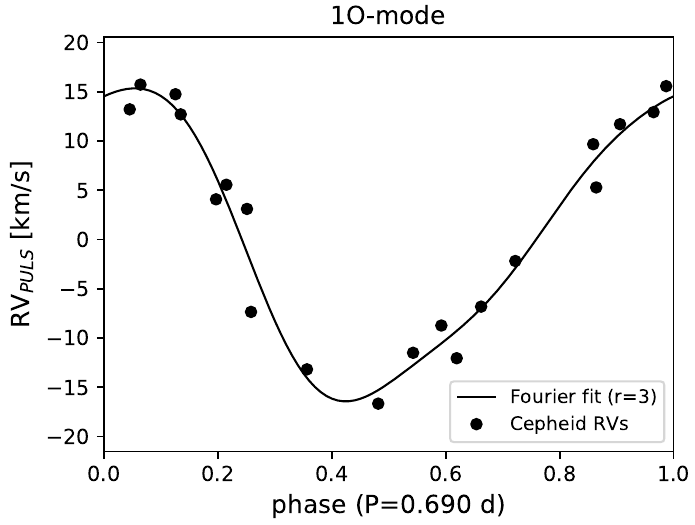}
    \end{center}
\caption{(left) Orbital radial velocity curve with the $1O$-mode pulsation of the Cepheid subtracted. The orbit is circular, and the amplitude ratio indicates the companion is about 2 times less massive. (right) Pulsational RV curve for $1O$ mode with orbital motion subtracted. The scatter of the Cepheid RVs around the fits comes mostly from the unaccounted $2O$ pulsation. }
\label{fig:1347}
\end{figure}

One of the most interesting objects in the sample is the binary double-mode Cepheid OGLE-LMC-CEP-1347. It has an orbital period of only 58.6 days, about five times shorter than the shortest known before for a binary Cepheid \citep{cep1347_ApJL_2022}. The Cepheid itself is the shortest-period one ever found in a binary system and the first double-mode Cepheid in a spectroscopically double-lined binary.
OGLE-LMC-CEP-1347 is most probably on its first crossing through the instability strip, as inferred from both its short period and fast period increase, consistent with evolutionary models, and from the short orbital period (not expected for binary Cepheids whose components have passed through the red giant phase). Our evolutionary analysis yielded a first-crossing Cepheid with a mass between 3 and 3.5 M$_\odot$ (lower than any measured Cepheid mass). The companion is a stable star, at least two times fainter and less massive than the Cepheid (mass ratio $q$=0.56), while also redder and thus at the subgiant or more advanced evolutionary stage. To match these characteristics, the Cepheid has to be a product of binary interaction, most likely a merger of two less massive stars, which makes it the second known classical Cepheid of binary origin. The orbital and pulsational radial velocity curves are shown in Fig.~\ref{fig:1347}

%%%%%%%%%%%%%%%%%%%%%%%%%%%%%%%%%%%%%%%%%%%%%%%%%%%%%%%%%%%%%%%%%%%%%%%%%%%%%
\acknowledgements
% Do not leave a blank line here! <---------------------->
The research that led to the presented results received funding from the Polish National Science Center grant SONATA BIS 2020/38/E/ST9/00486.

\bibliography{C02_pilecki}
\bibliographystyle{caosp309}

\end{document}